\documentstyle[aas2pp4]{article}

\input{psfig}
\input{psfig.sty}

\begin{document}

\title{SUNSPOT TRANSITION REGION OSCILLATIONS\\ 
IN NOAA 8156}

\author{N. Brynildsen, T. Leifsen, O. Kjeldseth-Moe, P. Maltby} 
\affil{Institute of Theoretical Astrophysics, University of Oslo, P.O. Box 1029 Blindern, \\
0315 Oslo, Norway}

\and

\author{K. Wilhelm}
\affil{Max-Planck-Institut f\"ur Aeronomie, D-37191 Katlenburg-Lindau, Germany}

\begin{abstract}

Based on observations obtained with the {\it Solar and Heliospheric
Observatory - SOHO} joint observing program for velocity fields in
sunspot regions, we have detected 3~min transition region umbral
oscillations in NOAA 8156.
Simultaneous recordings of O~{\scriptsize V} $\lambda$629 and 
N~{\scriptsize V} $\lambda$1238, $\lambda$1242 with the SUMER instrument give 
the spatial distribution of power in the 3~min oscillations, both in 
intensity and line-of-sight velocity.
Comparing loci with the same phase we find that the entire umbral 
transition region oscillates.
The observed maxima in peak line intensity are nearly in 
phase with the maxima in velocity directed towards the observer.  
We discuss the suggestion that the waves are upward propagating 
acoustic waves.  

\end{abstract}


{\underline{To appear in: {\it Astrophysical Journal Letters}}}

\section{Introduction}

The study of oscillations in the atmosphere of sunspots dates
back to the detection of umbral flashes in the K line (Beckers 
\& Tallant 1969), followed by the detection in H$_\alpha$ 
of penumbral waves (Zirin \& Stein 1972, Giovanelli 1972) and 
umbral oscillations (Giovanelli 1972).
The nature and mode identification of the 3~min
oscillations in the umbral chromosphere have been debated.
Based on linear theory two cavities have been discussed,
{\frenchspacing i.e.} a cavity for fast modes in the low
photosphere and sub-photosphere and a cavity for resonant slow
modes in the umbral chromosphere as well as a unified model with
contributions from both cavities ({\frenchspacing e.g.} 
Thomas \& Weiss 1992). 
However, the 3 minute umbral oscillations in the chromosphere show a 
non-linear character, suggesting the presence of upward propagating 
shock waves (Lites 1992, Bard \& Carlsson 1997).
Knowledge of the oscillations in the transition region originates from 
the study of eight sunspots in the C~{\scriptsize IV} $\lambda$1548 line 
by Gurman et al. (1982), see also Henze et al. (1984).
Interestingly, the observed waves with periods of 129 - 173~s, show no 
signs of shocks and appear to be upward propagating acoustic waves.
Recently, Rendtel et al. (1998) reported transition region intensity 
oscillations in the 2~min range in sunspot NOAA 7986.

We focus on transition region sunspot oscillations
in NOAA 8156, observed simultaneously in O~{\scriptsize V} $\lambda$629
and N~{\scriptsize V} $\lambda$1238, $\lambda$1242.
The observations were obtained on 1998 February 18 between 16:00 UT and 
21:07 UT with the Solar Ultraviolet Measurements of Emitted Radiation 
(SUMER; Wilhelm et al., 1995) instrument as part of a 
joint observing program on the {\it Solar and Heliospheric 
Observatory (SOHO)}.

\section{OBSERVATIONS}

The pointing of the SUMER slit is kept constant to 525$\arcsec$~W and 
332$\arcsec$~S off the solar disk center, i.e. at a heliocentric
distance of 40$^{\circ}$. The solar rotation moves the image of 
the sunspot, with umbral size 15$\arcsec$ and penumbral size 
50$\arcsec$, over the narrow slit, 0.3$\arcsec \times$ 120$\arcsec$.
From each spectrum 90$\arcsec \times$ 2.2 {\AA} spectral windows, 
centered on O~{\scriptsize V} $\lambda$629
and N~{\scriptsize V} $\lambda$1238, $\lambda$1242, are recorded with 
detector B.
With an exposure time of 15 s it takes 20 min to observe 80 consecutive 
spectra, called one raster, while the solar rotation moves the image
2.2$\arcsec$.
Figure~1 (bottom) shows the positions of the 12 rasters and 
the location of the sunspot, between 300$\arcsec$ and 
350$\arcsec$~S off disk center.
Between each raster, exposures with the rear slit camera give
the slit position in relation to the sunspot and 100 s 
exposures of 90$\arcsec \times$ 40~{\AA} spectra give the wavelength scale, 
from the wavelength positions of chromospheric lines.
The data reductions include corrections for the fixed pattern noise, 
defects of the detector and corrections for the geometrical distortions, 
see Wilhelm et al. (1995).
A least squares fit of a single Gaussian to each observed line profile 
gives the peak line intensity, $I$, the relative line-of-sight velocity, 
$v$, and the line width, $w = FWHM/(2\sqrt{ln~2})$.
We estimate the accuracy in the relative line-of-sight velocity determinations
to be 1 km~s$^{-1}$ for O~{\scriptsize V} $\lambda$629 and 
1.5 km~s$^{-1}$ (2 km~s$^{-1}$) for N~{\scriptsize V} $\lambda$1238
($\lambda$1242).

\begin{figure}[htb] 
   \centerline{\psfig{figure=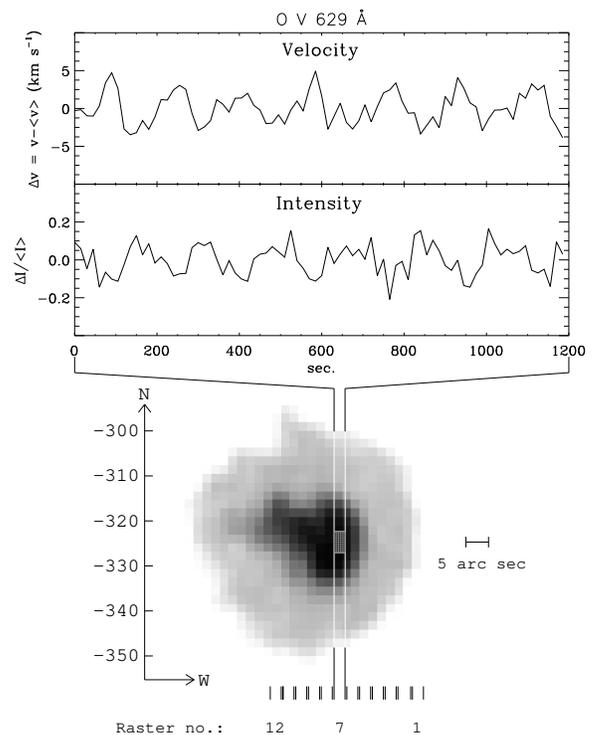,width=8cm}}
 \figcaption[]{Observed O~{\scriptsize V} $\lambda$629  
oscillations in relative line-of-sight velocity, $\Delta v = v - <v>$, and
relative peak intensity, $\Delta I/<I>$, in the center of the umbra, see
hatched area in the sunspot image, observed with 
the MDI instrument (Scherrer et al. 1995).}
\label{fig1}
\end{figure}

From the SUMER observations alone it is difficult to separate temporal
from spatial changes in oscillatory power.
Simultaneous observations with the Coronal Diagnostic Spectrometer 
(CDS; Harrison et al. 1995) in ten emission lines, 
including the O~{\scriptsize V} $\lambda$629 line, with a cadence of 25 min, 
show only small temporal variations in the overall intensity and 
line-of-sight velocity patterns during the time span of the observations. 
For a description of sunspot velocity field observations with CDS, 
see Brynildsen et al. (1998).  
The CDS results encourage us to use the SUMER observations to investigate 
the spatial distribution of the power in the transition region oscillations. 
By deriving the observed characteristics within each raster for 2$\arcsec$ 
sections along the slit the power distribution is studied within an area 
of 33.5$\arcsec$ $\times $ 90$\arcsec$ with a spatial resolution of 
2.2$\arcsec$ $\times$ 2$\arcsec$.

\section{Results}

As an example Figure~1 shows the observed O~{\scriptsize V} $\lambda$629 
umbral oscillations in relative line-of-sight velocity, 
$\Delta v~=~v~-~<v>~$, and relative peak line intensity, $\Delta I/<I>$.
The hatched area in the umbra marks the location of this oscillation.
Slow variations in the background fields have been
removed from the average values, $<v>$ and $<I>$.
This corresponds to applying a low frequency filter in
the Fourier domain.
Note that the observed oscillations are linear in character without
clear signs of shocks. 
The amplitudes of the oscillations are listed in Table~1.

\begin{table}[htb]
\caption{{\scriptsize OSCILLATION AMPLITUDES - SEE FIGURE~1}}
\begin{tabular}{lcccc}
\tableline
                 & Velocity        & Peak Intensity  &  & \\
Line             & (km s$^{-1}$)    & (\%)            &   & \\
\tableline
O~{\scriptsize V} $\lambda$629 &  2.7      & 11          &  &      \\
N~{\scriptsize V} $\lambda$1238 & 2.4      &  9          &  &      \\   
N~{\scriptsize V} $\lambda$1242 & 2.7      & 10          &  &      \\    
\tableline
\end{tabular}
\end{table}

\begin{figure}[htb] 
   \centerline{\psfig{figure=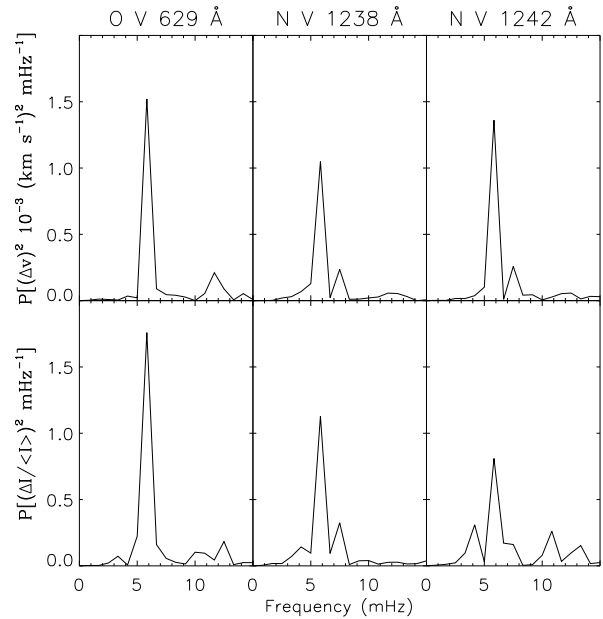,width=9cm}}
 \figcaption[]{Observed power in the transition region sunspot oscillation
in line-of-sight velocity and relative peak line intensity for the same 
position as in Figure~1.}
\label{fig2}
\end{figure}

Figure~2 shows the corresponding power spectra for the oscillations
in line-of-sight velocity and peak line intensity.
The power spectra are nearly identical with one maximum within the 
frequency interval 5.0~-~6.7 mHz, corresponding to periods between 150 and 
200~s, and maximum power close to 170 s. 
We also find a 5\% variation in the line width. The corresponding
power spectrum is similar to those shown in Figure~2, but is more noisy.
The results are scrutinized by tests where artificial noise is 
added to the observed signal and new values of the line parameters are 
derived.
The tests show that the results for the intensity and the line-of-sight 
velocity oscillations are confirmed, but the results are less certain 
for the line width.

Figure~1 shows near temporal coincidence of maxima in peak line intensity 
and maxima in blueshift.
From the corresponding power spectra we deduce a phase difference 
$\phi_v(629)$ - $\phi_I(629)$ = 173$^{\circ}$ between the line-of-sight 
velocity and the peak line intensity for the O~{\scriptsize V} line. 
The corresponding phase difference for the N~{\scriptsize V} line, 
$\phi_v(1238)$ - $\phi_I(1238)$ = 172$^{\circ}$.
We give less weight to the observed phase difference of 160$^{\circ}$
in the $\lambda$1242 line since this line is weaker than the other lines.
In Figure~5, below, we present this phase difference for the  
O~{\scriptsize V} $\lambda$629 line for the entire sunspot region.
 
The determination of the phase difference in velocity or intensity between 
the N~{\scriptsize V} $\lambda$1238 and O~{\scriptsize V} $\lambda$629 lines
is difficult since it involves careful alignment along the slit of lines 
observed in different locations on the detector.
For the umbra we find that the wave arrives on average 10~s~$\pm$~5~s 
earlier in N~{\scriptsize V} $\lambda$1238 line than in 
O~{\scriptsize V} $\lambda$629 line.  
If the N~{\scriptsize V} line originates at a lower height than the
O~{\scriptsize V} line, as seems reasonable considering the
temperatures of formation for the N~{\scriptsize V} (1.5 $\times$ 10$^5$K) 
and O~{\scriptsize V} (2.1 $\times$ 10$^5$K) lines, this suggests an 
upward propagating wave.

\begin{figure}[htb] 
   \centerline{\psfig{figure=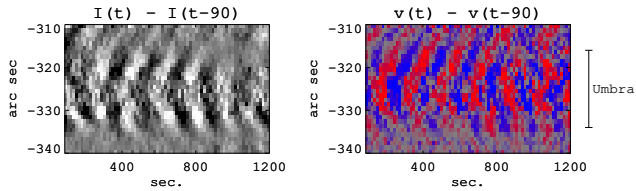,width=8cm}}
 \figcaption[]{Observed O~{\scriptsize V} $\lambda$629 variation 
in relative peak line intensity (left) and line-of-sight velocity 
(right) along the slit in raster 7. Blue (red) color corresponds to
motion towards (away from) the observer. 
To enhance the visibility, the observed signal at time $t$ is subtracted
from the observed signal at time $t$ -~90~s - this time difference
is close to half the wave period, 170~s.}
\label{fig3}
\end{figure}

Consider next the observed variation along the slit in raster 7,
{\frenchspacing i.e.} in the 2.2$\arcsec \times 90\arcsec$ wide raster 
that includes the central part of the umbra. 
Figure~3 gives the O~{\scriptsize V} $\lambda$629 variations
in peak line intensity, $\Delta I/<I>$, and relative line-of-sight velocity, 
$v - <v>$, as a function of time and position along the slit. 
The visibility and shape of the oscillation wave front is enhanced by 
subtracting the observed signal at time $t$ from the signal observed at 
$t$ - 90~s, a time difference close to half of the dominant period.
Figure~3 shows that waves with periods close to 170~s are observed
mainly within the umbra and do not extend into the penumbra to 
any great extent. 
The wave front in the central part of the umbra leads the wave 
front at the rim of the umbra by nearly a full wave period.
Clearly the observed oscillation affects the entire umbral
transition region, but the generation or transmission of
the wave depend on the position within the umbra. 

\begin{figure}[htb] 
   \centerline{\psfig{figure=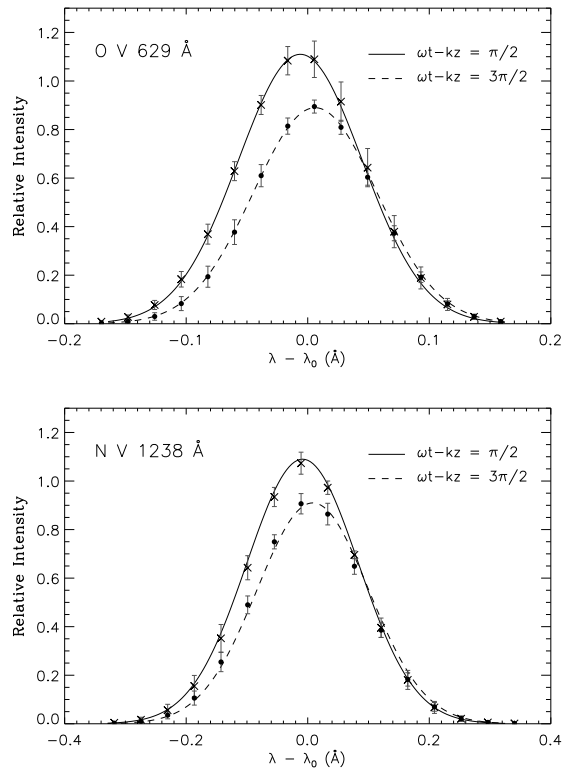,width=8cm}}
 \figcaption[]{Comparison of calculated line profiles that are formed 
in a region pervaded by upward propagating acoustic waves and 
observed values (dots and crosses), for ($\omega t - k z$) = $\pi/2$ 
and $3\pi/2$. The r.m.s. deviations between six observed values with 
the same phase are shown. The sunspot is located at $\theta$ = 40$^\circ$.}
\label{fig4}
\end{figure}

\begin{figure*}[htb] 
   \centerline{\psfig{figure=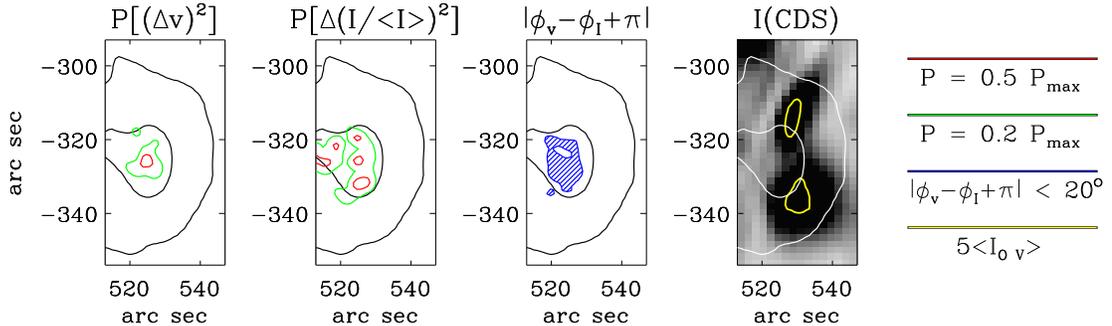,width=14cm}}
 \figcaption[]{Left to right: Spatial distribution of the 
O~{\scriptsize V} $\lambda$629 power in relative line-of-sight velocity, 
relative peak line intensity, phase difference $|\phi_v - \phi_I + \pi|$,
and the CDS peak line intensity image, where sunspot plumes 
with $I \ge 5 \times <I>$ are marked with yellow contours.
The contours of the umbra and penumbra are shown.}
\label{fig5}
\end{figure*}

To investigate the suggestion by Gurman et al. (1982) that the
oscillations are caused by upward propagating acoustic waves, we make
use of the property that the relative perturbation in number density,
$\Delta N/ N_{0}$, varies in phase with the line-of-sight velocity,
$v_{\parallel}$, ({\frenchspacing e.g.} Hansteen \& Maltby 1992).  
Regarding $v_{\parallel}$ as positive when
directed away from the observer we may write,

\begin{eqnarray}
{{{v_{\parallel,m}}\over {v_{S}}} sin (\omega t - k z)} = {-\,{{cos \theta}} {{\Delta N_{m}}\over {N_{0}}} sin (\omega t - k z)} 
\end{eqnarray}

\noindent 
for an upward propagating acoustic wave with a sinusoidal variation.  
Here $v_{S}$ is the sound speed, $\theta$ is the aspect
angle and the index $m$ denotes the amplitude. 
For N~{\scriptsize V} and O~{\scriptsize V}, formed in a gas with a 
mean molecular weight of 0.60 at temperatures close to 
1.5 $\times$ 10$^5$ K and 2.1 $\times$ 10$^5$ K, we estimate the sound 
speeds to be 58~km~s$^{-1}$ and 69~km~s$^{-1}$. 
We will use equation (1) to calculate line profiles that may be compared 
with observations. Since the lines may be
regarded as optically thin, the intensity is determined by an
integration along the line-of-sight of the product of the line
profile, the electron number density, $N_e$, the N~{\scriptsize V}
(O~{\scriptsize V}) ion density, $N_i$, and a temperature dependent
function, $g (T_e)$ ({\frenchspacing e.g.} Mariska 1992).  
The influence of the wave on the function $g (T_e)$ is small 
({\frenchspacing e.g.} Gurman et al. 1982).  
Assuming ionization balance, such that $N_i \propto N_e$, and 
neglecting possible variations in the thickness of the transition region, 
we may write the line intensity as 
$I_{\nu}(\theta) \propto N_e^2 \propto N^2$. This implies that we may
use the observed intensity amplitudes of 9\% and 11\%, see Table~1, to
determine the corresponding density amplitudes and equation (1) to
calculate the line profile for different values of ($\omega t - k z$).
Figure~4 shows the calculated line profiles for O~{\scriptsize V}
$\lambda$629 and N~{\scriptsize V} $\lambda$1238, with $\Delta N_{m} /
N_{0}$ = 0.055 and $\Delta N_{m} / N_{0}$ = 0.045, and line broadening owing 
to the instrument and non-thermal velocities of 27 km~s$^{-1}$ and 24
km~s$^{-1}$, respectively.  For ($\omega t - k z$) = $\pi/2$ both the
peak line intensity and the line shift towards shorter wavelengths
reach their maxima, whereas for ($\omega t - k z$) = $3\pi/2$ a maximum
in line shift towards longer wavelengths occurs while the peak line
intensity reaches a minimum.

The observed values at each wavelength are averages of six intensity
values from the center of the umbra at the corresponding phases.  
The averages are determined from the time series in raster 7,
where six minima and maxima are observed.
The average is constructed from a 2$\arcsec$ wide strip centered 
at position -326$\arcsec$, see Figure~3.  
The agreement between observed and calculated profiles in
Figure~4 is remarkable and supports the interpretation of the
oscillations as caused by upward propagating acoustic waves.
To obtain a deeper understanding of this result, the wave propagation
through the transition region should be thoroughly discussed, see 
Zugzda, Staude, \& Locans (1984) for a discussion of this topic. 

Figure~5 gives the spatial distributions of power in  
line-of-sight velocity and peak line intensity oscillations 
between 5.0 - 6.7 mHz for the O~{\scriptsize V} $\lambda$629 line.
The phase difference $|\phi_v - \phi_I + \pi| < 20^\circ$
throughout most of the umbra, {\frenchspacing i.e.} the maximum 
intensity is nearly in phase with the maximum velocity directed 
towards the observer.  
The oscillations are restricted to the umbra and interestingly, the 
spatial distributions of power in velocity and intensity differ. 
Whereas the power in velocity shows one maximum, the power in 
intensity also show maxima closer to the umbral rim, not far from 
the sunspot plumes observed with CDS (Figure~5, right), the plumes have
peak line intensity $I \ge 5 \times <I>$.
For a recent discussion of sunspot plumes, see Brynildsen et al. (1998). 
Possibly the direction of wave propagation changes with position in
the umbra.

In summary, significant oscillations in line-of-sight velocity and 
peak line intensity are observed, with periods close to 170 s.
The observed transition region oscillations are concentrated to the umbra
and the entire umbral transition region oscillates.
We study the observed line profiles and find support for the suggestion 
that the waves are upward propagating acoustic waves.
Discussions of the relation between the shock waves observed
in the chromosphere and the present observations are of interest, but 
outside the scope of this letter.

\acknowledgements 

We are indebted to the international SUMER team and thankful to the
MDI team for permission to use their data.
SUMER is supported by DLR, CNES, NASA and the ESA Prodex programme
(Swiss contribution).  SOHO is a project of international cooperation
between ESA and NASA.



\end{document}